\documentclass[sigconf, nonacm]{acmart}
\usepackage{amsmath,amsfonts}
\usepackage{algorithm}
\usepackage{algorithmicx}
\usepackage{algpseudocode}
\usepackage{graphicx}
\usepackage{textcomp}
\usepackage{listings}
\usepackage{xcolor}
\usepackage[english]{babel}
\usepackage{color,colortbl}
\usepackage{xspace}
\usepackage{multirow}
\usepackage{graphicx}
\usepackage{booktabs}
\usepackage{url}
\usepackage{xurl}
\usepackage{tipa}
\usepackage[T1]{fontenc}
\usepackage[utf8]{inputenc}
\usepackage{threeparttable}
\usepackage{amsmath,amsfonts}
\usepackage{pifont}
\usepackage{tabularx}
\usepackage[printonlyused]{acronym}
\usepackage[most]{tcolorbox}

\usepackage{hyperref}
\hypersetup{
	pdftitle={},
	pdfauthor={},
	pdfsubject={},
	pdfkeywords={},
	bookmarksnumbered=true,
	bookmarksopen=false,
	colorlinks=true,
	pdfstartview=FitB,
	pdfpagemode=UseOutlines
}

\definecolor{background}{HTML}{EEEEEE}
\definecolor{mygreen}{rgb}{0,0.6,0}
\definecolor{mygray}{rgb}{0.5,0.5,0.5}
\definecolor{mymauve}{rgb}{0.58,0,0.82}

\lstdefinelanguage{json}{
    basicstyle=\footnotesize\ttfamily,
    commentstyle=\color{black}, 
    stringstyle=\color{black}, 
    numbers=left,
    numberstyle=\scriptsize,
    stepnumber=1,
    numbersep=8pt,
    showstringspaces=false,
    breaklines=true,
    frame=lines,
    backgroundcolor=\color{background}, 
    string=[s]{"}{"},
    comment=[l]{:\ "},
    morecomment=[l]{:"},
}

\newcommand{\libcirc}[1]{\ding{\numexpr171+#1\relax}}

\usepackage{tikz}
\usetikzlibrary{shapes,arrows,positioning,fit}
\usetikzlibrary{backgrounds}

\usepackage{caption}
\newcommand{\zhiyun}[1]{\textcolor{blue}{[Zhiyun: #1]}}

\newcommand{\yu}[1]{\textcolor{mygreen}{[Yu: #1]}}

\newcommand{\haonan}[1]{\mytodogrey{[haonan: #1]}}
\newcommand{\mytodogrey}[1]{\textcolor{cyan}{\ding{46}~{\sf}~#1}}

\newcommand{\etal}{\textit{et al.}\xspace}
\newcommand{\ie}{\textit{i.e.,}\xspace}    
\newcommand{\eg}{\textit{e.g.,}\xspace}    

\newcommand{\squishlist}{
	\begin{list}{$\bullet$} {
			\setlength{\itemsep}{0pt}
			\setlength{\parsep}{0pt}
			\setlength{\topsep}{0pt}
			\setlength{\partopsep}{0pt}
			\setlength{\leftmargin}{1.0em}
			\setlength{\labelwidth}{1em}
			\setlength{\labelsep}{0.5em}
		}
		}
		\newcommand{\squishend}{
	\end{list}
}

\newcommand{\find}[1]{
\begin{tcolorbox}[leftrule=1mm,rightrule=1mm,toprule=0mm,bottomrule=0mm,left=1pt,right=1pt,top=0.5pt,bottom=0.5pt]
\em #1
\end{tcolorbox}
}

\newcommand{\cut}[1] {}

\cut{\zhiyun{a few things we should do before submission:
		\begin{itemize}
			\item check captilization consistency, ``figure -> Figure'', caption of a figure/table should have the first letter captilized only
			\item center the caption of all figures/tables.
			\item remove the ``period'' mark in captions.
			\item spell check / grammar check (put all sentences in grammarly)
			\item for all the functions referened in textit\{\}, we should add the brackets. For example, ``textit\{fun\} -> textit\{fun()\}''
			\item global replace of ``\ textit'' with ''\ texttt''. Also, there is no need to add the quotation marks around \texttt.
			\item check the text that references the figure and the figure itself are always co-located on the same page (or one page before/after).
		\end{itemize}
	}}

\acmConference[SomeConf ]{xxt  Conference on Something}{Aug 2023}{Earth}

\AtBeginDocument{%
  }

\author{Haonan Li}
\email{hli333@ucr.edu}
\affiliation{%
    \institution{UC Riverside}
    \city{Riverside}
    \state{California}
    \country{USA}
}

\author{Yu Hao}
\email{yhao016@ucr.edu}

\affiliation{%
    \institution{UC Riverside}
    \city{Riverside}
    \state{California}
    \country{USA}
}

\author{Yizhuo Zhai}
\email{yzhai003@ucr.edu}

\affiliation{%
    \institution{UC Riverside}
    \city{Riverside}
    \state{California}
    \country{USA}
}

\author{Zhiyun Qian}
\email{zhiyunq@cs.ucr.edu}

\affiliation{%
    \institution{UC Riverside}
    \city{Riverside}
    \state{California}
    \country{USA}
}

\newcommand{\work}{\mbox{\textsc{LLift}}\xspace}
\newcommand{\PP}[1]{
	\vspace{2px}
	\noindent\textbf{#1}
}
\newcommand{\yizhuo}[1]{\textcolor{orange}{[Yizhuo: #1]}}
\begin{document}

\title{The Hitchhiker's Guide to Program Analysis: A Journey with Large Language Models}

\newacro{LoC}{\textit{lines of code}}
\newacro{LLM}{\textit{Large Language Model}}
\newacro{UBI}{\textit{Use Before Initialization}}
\newacro{KB}{\textit{Inherent Knowledge Boundaries}}
\newacro{PS}{\textit{Path Sensitivity}}
\newacro{TPS}{\textit{Tradeoff Between Precision and Scalability}}

\begin{abstract}

Static analysis is a widely used technique in software engineering for identifying and mitigating bugs. 
However, a significant hurdle lies in achieving a delicate balance between precision and scalability. 
\acp*{LLM} offer a promising alternative, as recent advances demonstrate remarkable capabilities in comprehending, generating, and even debugging code. Yet, the logic of bugs can be complex and require sophisticated reasoning and a large analysis scope spanning multiple functions. 
Therefore, at this point, LLMs are better used in an assistive role to complement static analysis. 
In this paper, we take a deep dive into the open space of LLM-assisted static analysis, using use-before-initialization (UBI) bugs as a case study. 
To this end, we develop \work, a fully automated framework that interfaces with both a static analysis tool and an LLM. By carefully designing the framework and the prompts, we are able to overcome a number of challenges, including bug-specific modeling, the large problem scope, the non-deterministic nature of LLMs, etc.
Tested in a real-world scenario analyzing nearly a thousand potential UBI bugs produced by static analysis, \work demonstrates a potent capability, showcasing a reasonable precision (50\%) and appears to have no missing bug. It even identified 13 previously unknown UBI bugs in the Linux kernel. 
This research paves the way for new opportunities and methodologies in using LLMs for bug discovery in extensive, real-world datasets. 

\end{abstract}

\settopmatter{printfolios=true}

\maketitle

\section{Introduction}


Static analysis is a popular technique in software engineering, particularly in the area of bug discovery, that can improve code quality, reliability, and security.
However, the effectiveness of these techniques is influenced by the fundamental trade-off between precision and scalability, especially when dealing with extensive and complex programs~\cite{DBLP:journals/csur/ParkLR22, gosain_static_2015}.
On the one hand, static analysis solutions with lower precision tend to generate numerous false positives. 
On the other hand, expensive static analysis or symbolic execution solutions with higher precision often struggle to complete the analysis. 
Consequently, achieving comprehensive and accurate static program analysis for sizable programs like the Linux kernel poses a significant challenge.

\cut{
However, despite its numerous advantages, static analysis is not without its limitations. As we delve deeper into the intricate dynamics of large-scale software systems, these limitations start to surface and can often impede the precise and comprehensive analysis of the system. In this paper, we discuss three primary constraints encountered in the realm of static analysis: \textit{inherent knowledge boundaries}, \textit{exhaustive path exploration}, and \textit{rigidity in rule expansion and sensitivity adjustment}. Each of these challenges poses significant hurdles in our pursuit of a more refined and accurate static analysis, particularly in the context of vast and intricate systems like the Linux kernel.
}

UBITect \cite{ubitect}, a powerful static analysis solution illustrates these inherent limitations thoroughly. Targeting Use-Before-Initialization (UBI) bugs in the Linux kernel, it packages a pipeline of (1) a scalable bottom-up summary-based static analysis with limited precision, and (2) a precise symbolic execution with limited scalability.
The solution illuminates the need for alternative strategies to navigate the complex trade-offs between precision and scalability effectively.
Despite this strategic combination of analysis techniques, nearly 40\% of the potential bugs reported from the static analysis phase experience a timeout or memory exhaustion during the static symbolic execution phase, preventing any conclusive results on such cases. 
This limitation hinders the overall effectiveness of the tool, leading to the potential of two distinct outcomes: \textit{missed bugs} if these potential bug reports are ignored (what UBITect performs), or \textit{false positives} if they are sent to developers for inspection.

In this paper, we investigate the possibility of leveraging \acfp{LLM} 
as an alternative to handle such ``difficult cases''. This is because recent LLMs have exhibited strong potential in understanding, generating, and even debugging code ~\cite{github_github_nodate, langchain_2023_announcing_2023, codex}. 
Nevertheless, navigating the intricacies of utilizing LLMs for bug discovery proves to be a complex feat.
The technical report on GPT-4 underscores this challenge, admitting that when it comes to discovering new vulnerabilities, it may not be the best solution standalone \cite{openai_2023_gpt_4}: ``... is less effective than existing tools for complex and high-level activities like
novel vulnerability identification''.
In the same vein, prior research demonstrates the competence of LLMs mostly in simpler tasks or programs~\cite{pei_can_2023, ahmed_improving_2023, pearce_examining_2023}.
This is because LLMs are far from perfect. For instance, they suffer from \textit{hallucination}~\cite{ji_survey_2023} where instead of identifying the bugs in faulty code, LLMs may create non-existent facts in an attempt to rationalize the original intention behind the problematic code \cite{ma_scope_2023, tian_is_2023}. 
Another issue is the stochasticity of LLMs which can result in inconsistent or outright incorrect results, thus throwing another wrench into the gears of bug discovery~\cite{zhao2023survey}. Finally, LLMs have limited context windows, meaning they can only scrutinize a relatively small codebase.

In response, we propose \work, a fully automated framework that bridges static analysis with LLMs in analyzing UBI bugs.
Our solution packages several novel components.
First, \work performs \textit{post-constraint guided path analysis}, which helps verify the path feasibility of the ``use'' of an initialized variable, a difficult task for static analysis and symbolic execution.
Second, to efficiently interact with LLMs, we employ \textit{task decomposition} to break down the analysis into more than a single step. 
Third, we employ \textit{progressive prompting} by providing information incrementally only when necessary, instead of providing an enormous scope of code at once.
Finally, we propose \textit{self-validation} by requesting LLMs to review responses at various stages to obtain accurate and reliable responses. 

We implement a prototype of \work and test it in real-world scenarios. 
Focusing on the inconclusive cases of UBITect caused by time or memory limitation, \work successfully identifies 13 previously unknown UBI bugs in the Linux kernel that we confirmed with the Linux community. 
With 26 positive reports out of nearly 1,000 cases, \work reaches a high precision of 50\%. 
We also test \work against all previously known bugs found by UBITect, and observe a recall of 100\%.

We summarize our contributions as follows: 
\squishlist
\item \textbf{New Opportunities.} We introduce a novel approach to static analysis that enhances its precision and scalability at the same time by harnessing the capabilities of LLMs. 
To the best of our knowledge, we are the first to use LLMs to assist static analysis in bug-finding tasks with large-scale and real-world datasets.
\item \textbf{New Methodologies.} We develop \work, an innovative and fully automated framework that arms static analysis with LLMs. \work employs several prompt strategies to engage with LLMs, eliciting accurate and reliable responses.
\item \textbf{Results.} We rigorously investigate \work by conducting an in-depth analysis of nearly 1000 cases, resulting in a reasonable precision rate (50\%). Additionally, our examination led to the discovery of 13 previously unknown bugs.
\item \textbf{Open source.} Committed to open research, we will publicly release all of our code and data, fostering further exploration of the new space of LLM-assisted program analysis.
\squishend

\begin{figure}[]
\hspace{-15pt}
\includegraphics[width=.5\textwidth]{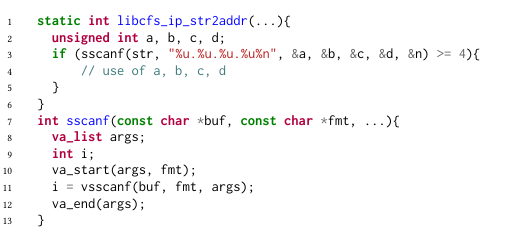}
\caption{Code snippet of \texttt{sscanf} and its usecase}
\label{fig:sscanf}
\end{figure}

\begin{table}[]
  \caption{UBITect's summary  for \texttt{sscanf}. Both \textit{use} and \textit{initialization} for \texttt{va\_args} are incorrect.
  \ding{51} and \ding{55} stand for whether this parameter will be used/initialized after its call.
   ``...'' represents all other parameters of \texttt{va\_args}.} 
  \begin{tabular}{l|ccccc}
  \toprule
   & \texttt{buf} & \texttt{fmt} & ... & \texttt{*buf} & \texttt{*fmt}  \\ \midrule
  Use & \ding{51} & \ding{51} & \ding{51} &  \ding{51} & \ding{51}  \\
  Initialize & \ding{55} & \ding{55} & \ding{55} & \ding{55} & \ding{55}\\ \bottomrule
  \end{tabular}
  \label{tab:sscanf_sum}
\end{table}

\section{Background \& Motivation}
\label{sec:moti}


\subsection{UBITect and Motivating Example} 
\label{sec:ubitect}

UBITect is a state-of-the-art static analysis solution aiming at finding \ac{UBI} bugs in the Linux kernel~\cite{ubitect}. 
It employs a two-stage pipeline where the first stage employs a bottom-up summary-based static analysis of the Linux kernel. 
By design, this stage aims for scalability and sacrifices precision, producing a significant number of potential bugs (\ie $\sim$140k), most of which are false alarms. 
The static analysis is imprecise partly due to its lack of path sensitivity (often needed to discover UBI bugs). 
It is complemented by a second stage of static symbolic execution that filters as many false alarms as possible by verifying their path feasibility.
However, 40\% of the reported bugs are discarded due to timeout (10 minutes) or memory limitations (2 GB) during the symbolic execution, potentially missing genuine bugs.



Figure \ref{fig:sscanf} shows a case where UBITect's static analysis stage considers it a potential UBI bug (a false alarm) and the subsequent symbolic execution stage times out and fails to generate a definitive conclusion. In other words, UBITect failed to rule out this case as a false alarm. 
As Table \ref{tab:sscanf_sum} presents, the static analysis stage generates a summary of \texttt{sscanf()} as ``\textit{may not initialize}
parameters \texttt{a}, \texttt{b}, \texttt{c}, and \texttt{d}'' but \textit{does use them} at Line 3.
Consequently, the static analysis stage reports two locations of \textit{use-before-initialization} at Line 3 and Line 4, respectively.
There are two reasons for the static analysis stage to consider the case a potential bug: 
1) \textbf{\textit{inability to recognize special functions}}: 
For soundness, UBITect assumed the \texttt{va\_start()} is a normal function. However, since it cannot find its definition, it has to conservatively assume that the arguments passed to it will be used inside.
Unfortunately, in reality, \texttt{va\_start} is a compiler built-in function that simply ``prepares' the arguments without any uses.
2) \textbf{\textit{insensitivity of path constraints}}: 
It fails to recognize the path constraint, \ie \texttt{if(sscanf(...)>=4)}, which ensures its arguments \texttt{a} to \texttt{d} must be initialized before use.

\subsection{Practical Challenges of Static Analysis}
\label{subsec:funda_chall}


In light of our motivating example of the \texttt{sscanf()} case, we 
can summarize the reasons for UBITect's failure as follows:

\vspace{3pt}
\noindent \textbf{Inherent Knowledge Boundaries.} 
Developers need to model specific functions or language features. Otherwise, they influence the correctness of the results. 
For compiler built-in functions, \eg \texttt{va\_start()}, their definitions are simply not available. 
Beyond this example, there exists an array of other scenarios, which are particularly prevalent in the Linux kernel. These situations include assembly code, hardware behaviors, callback functions, concurrency, and compiler built-in functions. However, in practical terms, it is often time-consuming to discover and model all these cases, because they can be highly dependent on the analysis target and evolve over time. This limitation often compromises the effectiveness of static analysis, leaving it less precise and comprehensive than desired. 


\cut{
\vspace{3pt}
\noindent
\textbf{Lack of adaptability in Rule Expansion and Sensitivity Adjustment.}
Traditional static analysis tools rely on pre-defined rules that are often fixed and difficult to modify. If new rules or patterns need to be added or existing ones modified, it requires manual intervention and expertise in the underlying codebase. This lack of flexibility makes it challenging to adapt the analysis to evolving codebases or handle complex scenarios.
Even though UBITect was specifically designed to discover UBI bugs in the Linux kernel, it falls short of being aware of the postconditions to rule out false alarms.
Unfortunately, tailoring the degree of
analysis sensitivity is no less challenging, \eg it may strain the tool's scalability with increased sensitivity and add complexity to the analysis process.
}

\vspace{3pt}
\noindent \textbf{Exhaustive Path Exploration.}
Correctly handling cases like \texttt{sscanf()} requires it to consider the check: \texttt{sscanf(...)>=4}. Unfortunately, 
existing path-sensitive static analysis (and symbolic execution) techniques operate under a methodical but exhaustive paradigm, exploring all potential execution paths through the codebase. While this approach is theoretically comprehensive, it often leads to a combinatorial explosion. The vast array of execution paths necessitates the exploration of myriad functions, many of which ultimately prove irrelevant to the specific analysis task at hand. 
In the \texttt{sscanf()} case, its return value is computed inside an unbounded loop when iterating over an unknown string variable \texttt{buf}.
This causes UBITect's symbolic execution to time out exactly due to this problem.

\cut{
Static analysis techniques typically exhibit a certain level of inflexibility regarding
rule extension and context sensitivity adjustments. Adding new rules or
heuristics often means delving into the depths of the tool's codebase, requiring
expert intervention and a robust testing regime --- a process that can be
time-consuming and susceptible to error. 
Meanwhile, tailoring the degree of
context sensitivity is no less challenging, demanding a profound understanding
of the underlying analysis algorithms and potential ripple effects on the
results. Moreover, incorporating new rules or increased context
sensitivity may strain the tool's scalability, adding complexity to the
analysis process. Thus, broadening the rule set or modifying the
context sensitivity embodies a significant challenge for conventional static
analysis.
}




\subsection{Capability of LLMs}
\label{subsec:cap}

Fortunately, \acp*{LLM} \cite{openai_2023_gpt_4} offers a promising
alternative to summarizing code behaviors~\cite{openai_training_2022} in a flexible way and bypassing the aforementioned challenges. This is because LLMs
are trained and aligned with extensive datasets that include both natural language and programs. 
Specifically, we observe that LLMs possess fundamental abilities that assist in addressing each challenge: 1) \textit{\textbf{domain-specific code recognition}} and 2) 
\textit{\textbf{smart code summarization}}.

\vspace{3pt}
\noindent
\textbf{Domain-specific Programming Constructs Recognition.} 
This proficiency is showcased in three key areas: 1) \textit{\textbf{Function Recognition}}: LLMs can identify frequently used interfaces in the Linux kernel
from its semantics, such as \texttt{sscanf()}, \texttt{kzalloc()}, \texttt{kstrtoul()}, and  \textit{`list for each'}, simplifying the analysis and making the analysis more scalable.
2) \textit{\textbf{Function pointers and callbacks}}: LLMs can accurately interpret complex uses of function pointers as callbacks, which often require manual modeling. We will show an interesting case in \S\ref{subsec:case_study}.

\cut{
LLMs display an impressive ability to
understand and interpret various advanced programming constructs that are specific to a domain, e.g., Linux kernel. In our own analysis, we observed LLMs are capable of recognizing several types of code patterns: 1) \textbf{\textit{Inline Assembly:}}
LLMs understand assembly instructions, helping to bridge the gap
between high-level code and machine instructions. 
This can help with summarizing the behavior of a function (which embeds inline assembly) more precisely.
2) \textit{\textbf{Function Pointers and
Callbacks:}} LLMs, like ChatGPT, have the capability to accurately interpret the
complex use of function pointers as callbacks, \eg specific callback APIs. This ability allows them to
anticipate potential impacts on the program state, thus enabling a broader and
more accurate analysis that can be difficult for static methods.
3) \textit{\textbf{Common APIs:}} LLMs recognize frequently used APIs and programming constructs in the Linux kernel, such as `list for each', \texttt{kzmalloc}, \texttt{kstrtoul} and
\texttt{get\_user\_pages\_unlocked}.
}

\vspace{3pt}
\noindent
\textbf{Smart Code Summarization.} 
LLMs can work with complicated functions;
for example, that they can summarize loop invariants \cite{pei_can_2023}, which is an inherently difficult task in program analysis. 
This is likely because it has been trained on various functions with loops and their semantics.
In contrast, traditional static analysis follows explicitly defined rules 
without a limited ability to generalize.

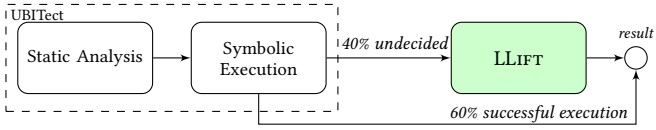
\begin{figure}
  \centering
  \scalebox{1.0}{
  \hspace{-5pt}
  \begin{tikzpicture}[
    block/.style={rectangle, draw, text width=5em, text centered, rounded corners, minimum height=3em},
    line/.style={draw, -latex'},
    node distance=0.3cm and 0.5cm
]

\node[block, font = \footnotesize] (static) {Static Analysis};
\node[block, font = \footnotesize, right=of static] (symbolic) {Symbolic Execution};

\path[line] (static.east) -- (symbolic.west);

\node[block, fill = green!20, right=1.65cm of symbolic, font = \small] (chatgpt) {\work};

\node[circle, draw,  right= of chatgpt, text centered, minimum height=0.1pt, label=\textit{\scriptsize result}] (final) {};

\path[line] (symbolic.east) -- (chatgpt.west) node [pos =0.55, sloped, above, font=\footnotesize] {\textit{40\% undecided}};
\path[line] (chatgpt.east) -- (final.west);

\node[dashed,  draw, fit={(static) (symbolic)}, inner sep=0.15cm, minimum height=4.5em] (ubitect_box) {};


\node[anchor=north west, inner sep=2pt, font=\scriptsize] at (ubitect_box.north west) {UBITect};

\path[line] (symbolic.south) |- ++(0,-0.4cm) -| (final.south) node[pos=0.6, above, left=0, font=\footnotesize] {\textit{60\% successful execution}};

\end{tikzpicture} 
  }
  \caption{The overview of \work. Start with the discarded cases by UBITect
   and determine whether these potential bugs are true or false.
   }
  \label{fig:design-flow}
  \vspace{-5pt}
\end{figure}

\begin{figure}[t]
\hspace{-15pt}
\includegraphics[width=.5\textwidth]{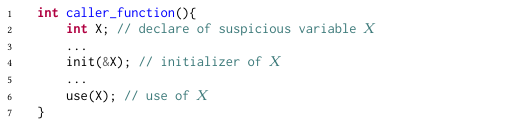}
\caption{A typical type of potential UBI bug. For each suspicious variable \(X\), we expect it to 1) have an initializer function that probably initializes  \(X\) and 2) use \(X\).  
}
\label{fig:prob_scope}
\end{figure}

\section{Problem Formulation}
\label{sec:problem}

\subsection{Definitions and Scope}

\subsubsection{Use-Before-Initialization.} A \acf{UBI} bug refers to the erroneous scenario where a variable \( v \) is accessed or involved in any operation prior to its correct initialization. 
Let:

\begin{itemize}
\item 
\(d(v)\) represent the declaration of \(v\).
\item 
\(u(v)\) signify a use operation involving \(v\).
\item 
\(i(v)\) denote the initialization operation of \(v\).
\end{itemize}

if there exists \(d(v)\) and \(u(v)\), then \(v\) is \textit{used before initialization} if:
\begin{equation}
\exists v : (d(v) < u(v)) \land \neg (\exists i(v) : d(v) < i(v) < u(v))
\label{eq:ubi_def}
\end{equation}
where \( < \) indicates a temporal sequence in the program execution. 

\subsubsection{Postcondition.}

Postconditions encapsulate the expected state or behavior of a system upon the conclusion of a routine~\cite{DBLP:books/ph/Meyer97}. Specifically, they detail the guarantees a routine offers based on its observable outcomes.

For a routine \( R \), consider its set of outcomes as
\(\mathcal{O}\). These outcomes are defined as \textit{updates} to its parameters (and return value) for a path of \(R\). Particularly, \(\mathcal{O}\) does not include
initialization for variables for convenience.
In the study of UBI bug, for a routine \( R \) that can yield a set of outcomes \( \mathcal{O} \), the postcondition \(\mathcal{P}\)
can be defined as:
\begin{equation}
\mathcal{P}_R: \mathcal{S}(R) \rightarrow  \mathcal{O} \times \texttt{must\_init}
\end{equation}
Here, \(\mathcal{S}(R)\) signifies all possible execution paths through the routine \(R\), 
\(\mathcal{O}\) describes all updates of \(R\) on its variables, and
\texttt{must\_init} is a set of variables that must be initialized. 

\PP{Motivating Example.}
Consider the \texttt{sscanf()} function in our motivating example. 
Based on these return values, the postconditions assure the initialization of certain variables:

\begin{align*}
    \mathcal{P}(path_1) &: \{{ret \mapsto 0}, \texttt{must\_init} \mapsto \emptyset \} \\
    \mathcal{P}(path_2) &: \{{ret \mapsto 1}, \texttt{must\_init} \mapsto \{a\} \} \\
    \mathcal{P}(path_3) &: \{{ret \mapsto 2}, \texttt{must\_init} \mapsto \{a,b\} \} \\
    \mathcal{P}(path_4) &: \{{ret \mapsto 3}, \texttt{must\_init} \mapsto \{a,b,c\} \} \\
    \mathcal{P}(path_5) &: \{{ret \mapsto 4}, \texttt{must\_init} \mapsto \{a,b,c,d\} \} \\
    \mathcal{P}(path_6) &: \{{ret \mapsto 5}, \texttt{must\_init} \mapsto \{a,b,c,d,n\} \} \\
\end{align*}

\noindent
Here, the \(path_1 - path_6\) represent different possible paths in the \texttt{sscanf()} and
each path corresponds with a different postcondition.

For UBI detection, not every associated postcondition is relevant; instead, only the outcomes making the \(u(v)\) reachable are \textit{critical}. The constraints of the use are  \textit{\textbf{post-constraints}} \( \mathcal{C}_{post} \) \cite{path_program_analysis}. The \textit{qualified postcondition}, \( \mathcal{P}_{qual} \), is a subset of \( \mathcal{P} \) refined by \( \mathcal{C}_{post} \):

\[
\mathcal{P}_{qual} = \mathcal{P} |_{\mathcal{C}_{post}}
\]

For the \texttt{sscanf()} case, if the post-constraint is \( \mathcal{C}_{post} = \text{ret} \ge 4 \), the qualified postcondition would be \(\mathcal{P}(path_5) \wedge \mathcal{P}({path_6})\), which ensures that variables \texttt{a, b, c,} and \texttt{d} must be initialized; therefore, all variables used subsequently are initialized, and no UBI happens.

In subsequent discussions, unless otherwise specified, the term \textit{`postcondition'} shall denote \textit{`qualified postcondition'}.

\subsection{Post-Constraint Guided Path Analysis}
\label{subsec:postcondi_work}

When analyzing a routine or function in a path-sensitive manner, the number of paths to explore can grow rapidly. Fortunately, if we have information about what the function is expected to achieve (given by \(\mathcal{C}_{post}\)), we can prune paths that inherently don't meet those expectations. We categorize
two scenarios, \textbf{\textit{direct application}} and \textbf{\textit{outcome conflicts}}, in applying this optimization.


Let \( R \) be the routine or function under analysis and \( \mathcal{S}(R) \) be its path set. Let \( path \in \mathcal{S}(R) \) refer to a specific path in \( R \). Besides, Each path \(path\) has an associated path constraint \(p\) that dictates its feasibility. These two optimizations can be formed with:


\PP{Direct Application.} For direct application, the post-constraint \(\mathcal{C}_{post}\) can be directly applied as a path constraint. 
A \textit{path} can be discarded if:

\begin{equation*}
\neg ( p(path) \land \mathcal{C}_{post}) 
\end{equation*}

This implies that if a \( path \) inherently contradicts the post-constraint, it can be removed from consideration.

\PP{Outcome Conflicts.} Let \( \mathcal{O}(p) \) denote the set of all outcomes or effects produced by path \( p \). A \textit{path} can be pruned if any of its outcomes conflict with the post-constraint:

\[
 \exists o \in \mathcal{O}(path) : \neg (o \land \mathcal{C}_{post})
\]

 This stipulates that if an outcome from \( path \) inherently contradicts the post-constraint, that path can be disregarded in the analysis.


\PP{Correctness.} The validity of these optimization methods can be proved by contradiction. Consider an instance where one of these paths is executed. If this path conflicts with the $\mathcal{C}_{\text{post}}$, it would render $u(v)$ unreachable. Thus, it becomes evident that such paths can be pruned without sacrificing the correctness of the analysis.

We provide a concrete example of how we perform these optimizations in \S\ref{subsubsec:postcon_rule}.



\cut{
\subsection{Assumptions}
\haonan{need to revise}
Let \(V\) be a variable, \(F\) be a function, and \(C\) be a callee function.

\begin{enumerate}
    \item \(A_1\): We restrict our analysis to patterns where \(V\) is declared in \(F\) and is passed to \(C\) for initialization.
    \item \(A_2\): Initialization of \(V\) in \(C\) can occur with multiple layers of indirection.
    \item \(A_3\): Direct initializations of \(V\) in \(F\) are excluded from the scope.
\end{enumerate}
}

\subsection{Conceptual Workflow}
\label{subsec:concept_wf}

Given a bug report containing a suspicious variable \( v \) and its residing function \( F \), the workflow \( \Phi \) is as follows:

\begin{enumerate}
    \item \( \Phi_1(F, v) \rightarrow \{i(v)\} \): Identify potential initializers for \( v \) from the bug report.
    \item \( \Phi_2(F, i(v)) \rightarrow \mathcal{C}_{post} \): Extract the \( \mathcal{C}_{post} \) from the bug report for each \(i(v)\).
    \item \( \Phi_3(F, \{i(v), \mathcal{C}_{post} \}) \rightarrow \text{InitStatus}(v) \): Summarize the initialization status for variable \( v \) after all possible initializers completion (merge multiple initializers). 
\end{enumerate}


\noindent
\textbf{Decision Policy.}
The decision policy \(\Delta\) is defined as: 
\begin{align*}
    \Delta(\text{InitStatus}(v) = \textit{must\_init}) & : \text{non-bug} \\
    \Delta(\text{InitStatus}(v) \neq \textit{must\_init}) & : \text{potential bug}
\end{align*}


In this policy, we adopt a conservative approach by treating all variables not explicitly marked as \textit{must\_init} as potential vulnerabilities.
And it is worth noting that this policy may introduce some false positives. For example, it might \textit{over-approximate} preconditions.

Conceptually, \work will not miss more bugs. The post-constraint guided path optimizations and decision policies are safe.

\subsection{Turns and Conversations in LLMs}
\label{subsec:turn_convo}


We define two key concepts in interacting with LLMs: \textit{turn} and \textit{conversation}.

\squishlist
    \item \textbf{Turn:} A turn encapsulates a singular interaction with the LLM. Formally, it's defined as a tuple, $(p, r)$, where $p$ represents the problem or question, 
    and $r$ denotes the LLM's response.

    \item \textbf{Conversation:} Leveraging the capabilities of LLMs often necessitates a series of interactions, especially for complex problem-solving. A conversation is an ordered sequence of turns. A conversation comprising $n$ turns can be expressed as $[(p_1, r_1), (p_2, r_2), \ldots , (p_n, r_n)]$.
\squishend

\section{Design} 
\label{sec:design}



\begin{figure}
    \hspace{-25pt}
    \includegraphics[width=.51\textwidth]{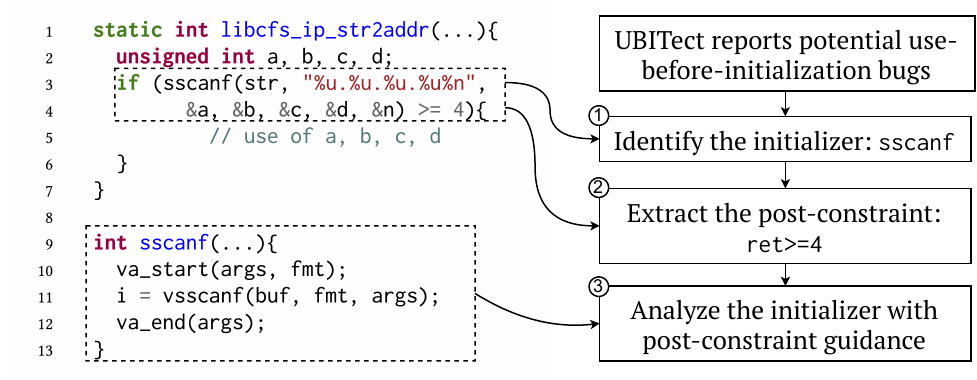}
\caption{Example run of \work. For each potential bug, \work \  \libcirc{1} (\(\Phi_{1}\)) identifies its initializer, 
    \libcirc{2} (\(\Phi_{2}\)) extracts the post-constraints of the initializer, and
     \libcirc{3} (\(\Phi_{3}\)) analyzes the behavior of the initializer with the post-constraints via LLM. }
     \label{fig:wf-case}
\end{figure}

In Section \S\ref{subsec:concept_wf}, we introduced a conceptual workflow. Elaborating on that foundation, Figure~\ref{fig:wf-case} showcases a compelling illustration of our methodological approach. Yet, translating this workflow into practice presents its challenges. 
Even with the advanced knowledge and analytical capabilities of cutting-edge LLMs,
achieving optimal results remains a challenge. Throughout the development of \work, we identified several obstacles and subsequently introduced four distinct design components to effectively address these challenges.

\subsection{Design Challenges}
\label{sec:design_chall}

It is non-trivial to prompt LLMs effectively~\cite{zhao2023survey, prompt_engineering}. We meet the following challenges and propose solutions correspondingly in designing \work.

\squishlist
\item \textbf{C1. Limited Understanding of Post-constraint.} 
Despite LLMs (\eg GPT-4) are able to comprehend the definition of post-constraint and apply them in simple scenarios, we found their capacity to utilize this knowledge in actual program analysis—such as summarizing function behavior in line with specific post-constraint —to be limited. This critical limitation often results in unpredictable and inconsistent outcomes.

\item \textbf{C2. Token Limitations.} It is known that \acp{LLM} have token limitations. For example, GPT-3.5 supports 16k tokens and GPT-4 supports 32k 
tokens \cite{openai_2023_function}. This means that we do not want to copy a large number of function bodies in our prompts to LLMs.

\item \textbf{C3. Unreliable and Inconsistent Response. }
LLMs are known to result in unreliable and inconsistent responses due to
\textit{hallucination} and \textit{stochasticity} \cite{zhao2023survey}.
Stochasticity refers to the inherent unpredictability in the model's outputs \cite{vaswani_attention_2017}; and the hallucination refers to LLMs generating nonsensical or unfaithful responses \cite{ji_survey_2023, zheng_why_2023}.
By design, the stochasticity can be mitigated with lower \textit{temperature}, a hyperparameter controlling the degree of randomness in outputs~\cite{salamone_what_temp_2021}; however, 
 reducing temperature may impair the model's exploring ability  \cite{xu_systematic_2022} and therefore 
 may miss \text{corner} cases that result in vulnerabilities.

\squishend

\begin{figure}
  \centering
  \includegraphics[width=0.48\textwidth]{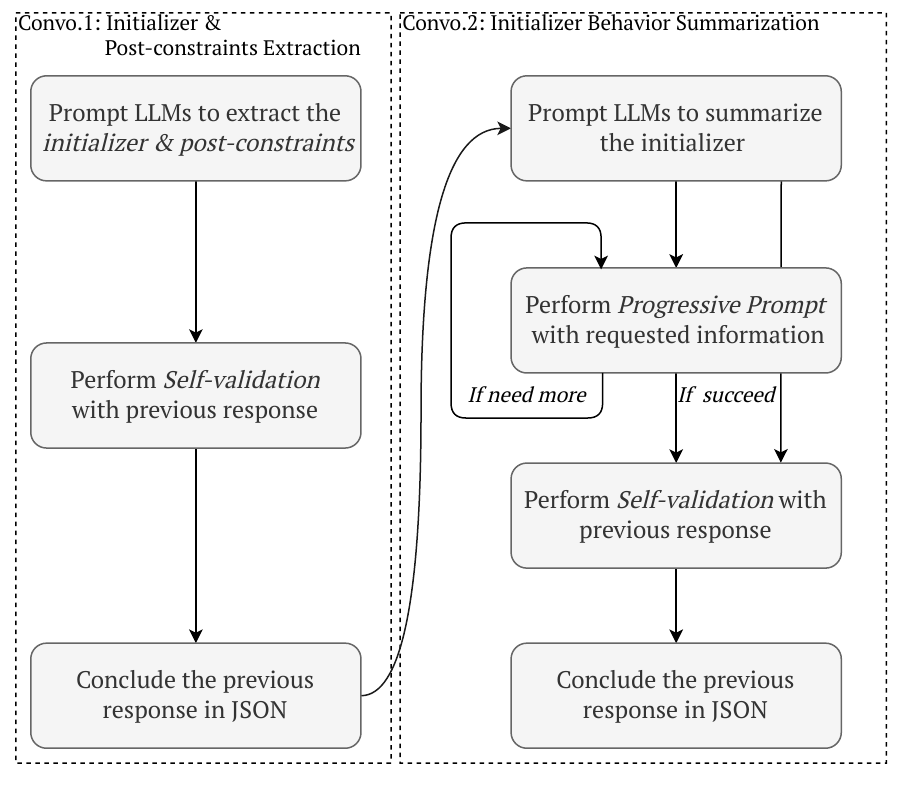}
  \caption{The workflow of \work. Given a potential bug, we let LLM first identify the initializer and then extract its post-constraints (Convo.1),  
  then leverage them to summarize the behavior of the initializer (Convo.2).
   A conversation consists of prompts (boxes) and responses (edges).} 
  \label{fig:wf}
\end{figure}

\subsection{Design Overview}

We will discuss our design strategies to address the above challenges in the rest of the section. Before that, we provide a high-level overview of our solution.

\squishlist
\item To tackle challenge \textbf{C1} (Post-constraint), we propose to encode \textbf{\textit{(D\#1) Post-Constraint Guided Path Analysis}} by teaching LLMs with examples, or \textit{few-shot in-context learning}, of post-constraints.
This approach enables LLMs to learn from a small number of demonstrative examples, assimilate the underlying patterns, and apply this understanding to process post-constraint guidance in our analysis. 

\item  To tackle challenge \textbf{C2} (Token Limitation), We employ two strategies: 
 \textbf{\textit{(D\#2) Progressive Prompt}.} 
Instead of copying a large number of function bodies (\ie subroutines), we only provide function details on demand, \ie when LLMs are not able to conduct a result immediately.
 \textbf{\textit{(D\#3) Task Decomposition.}} We break down the problem into sub-problems that can be solved in independent conversations, \ie \textit{a sequence of prompt and response pairs}. 

\item To tackle challenge \textbf{C3} (Unreliable Response), we employ the following strategies: 
\textbf{\textit{(D\#4) Self-Validation.}} We ask LLMs to review and correct their previous responses. This helps improve the consistency and accuracy based on our observation. 
Besides, \textit{\textbf{(D\#2) Progressive Prompt}} and \textit{\textbf{(D\#3) Task Decomposition}} also help to deal with this challenge. Additionally, we implement \textit{\textbf{majority voting}} by running each case multiple times and use majority voting to combat stochasticity. 
\squishend

We elaborate the design of (D\#1 - \#4) \textit{\textbf{Post Constraint Guided Path Analysis}}, \textbf{\textit{Progressive Prompts}}, 
\textbf{\textit{Task Decomposition}}, and
\textbf{\textit{Self-Validation}} detailed in the rest of this section.
The effectiveness and efficiency of these design strategies are rigorously evaluated in \S\ref{sec:comparison}, revealing a substantial enhancement in bug detection within the Linux kernel.

\subsection{Design \#1: Post-Constraint Guided Path Analysis}
\label{subsec:postcondi}

The Linux kernel frequently employs return value checks as illustrated in Table \ref{tab:postcondi_types}. Through our detailed examination of non-bug instances, we found that 
a path-sensitivity analysis can effectively eliminate over 70\% of these negative cases. However, path-sensitive static analysis usually suffers from path explosion, especially in large-scale codebases like the Linux kernel.

Fortunately, we can prompt the LLM to collect \(\mathcal{C}_{post}\) and summarize the function with respective to the 
\(\mathcal{C}_{post}\). It is worth noting that current LLMs (\eg GPT-4) are not natively sensitive to the sensitivity; without any additional instructions, LLMs usually overlook the post-constraints.
Therefore, we teach the LLM to be sensitive to post-constraints rules through few-shots in-context learning. We describe the design details as follows:

\begin{table}

\caption{Two types of post-constraints and their variants.}
\label{tab:postcondi_types}
\vspace{-3pt}
\includegraphics[width=.48\textwidth]{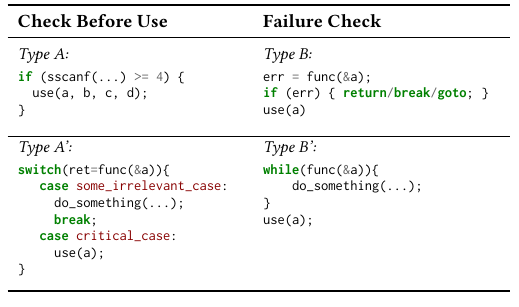}
\end{table}

\subsubsection{Post-Constraints Extraction.} 
To extract the \textit{qualified postcondition}, we first determine the post-constraints that lead to the use of suspicious variables.
We incorporate few-shot in-context learning to teach LLMs how to extract such constraints from the caller context. Table \ref{tab:postcondi_types} demonstrates how we teach LLM with in-context learning. We focus primarily on two types of code patterns:

\squishlist
\item \textbf{Check Before Use.} 
Type A is our motivating example; by looking at its check, the post-constraint should be \(ret \ge 4\).
Type A' describes a similar case with \texttt{switch-cases}, with expected output \(ret \mapsto \texttt{crticial\_case}\).
\item \textbf{Failure Check.} This pattern captures the opposite of the first pattern. They commonly occur in the Linux kernel where the error conditions cause the use to become unreachable, as illustrated in Type B, the post-constraint is \(err \mapsto 0\).
Type B' depicts a variant where the initializer keeps retrying til success, and therefore with expected output \(ret \mapsto 0\), which indicates
its first successful execution to break the endless loop.
\squishend

\subsubsection{Function Behavior Summarization.} 
Once we obtain the \textit{post-contraints} in Convo.1, we feed them to the LLM to obtain the behavior summary in Convo.2 . For example, we provide the following: 

\begin{lstlisting}[numbers=none]
{
 "initializer": "ret = sscanf(str,'%u.%u.%u.%u%n',&a,&b,&c,&d,&n)",
 "suspicious": ["a", "b", "c", "d"],
 "postconstraint": "ret >= 4"
}
\end{lstlisting}


The LLM may respond with 

\begin{lstlisting}[numbers=none]
{
 "ret": "success",
 "response": {
   "must_init": ["a", "b", "c", "d"],
   "may_init": [{"name":"n", "condition": "ret > 4"}]
  }
}
\end{lstlisting}

\cut{
We can see that the response encodes the postcondition concisely
where \texttt{a,b,c,d} are considered \texttt{must\_init} and \texttt{n}
is considered \texttt{may\_init} because it is initialized only when \(ret > 4\), and not when \(ret \mapsto 4\).
}

The response succinctly encapsulates the function behavior, where variables \texttt{a,b,c,d} are classified as \texttt{must\_init}, and \texttt{n} is categorized as \texttt{may\_init}. This is due to the initialization of \texttt{n} only occurring when \(ret > 4\), and not when \(ret \mapsto 4\).

Note that this seemingly simple interaction with LLMs can be challenging for static analysis or symbolic execution. Consider the \texttt{sscanf()} example, even if the analysis is aware that the qualified postcondition should be limited to those where \(ret \ge 4\),
it would still need to enumerate the paths inside of \texttt{sscanf()}, which involves loops and can easily lead to timeouts as explained in \S\ref{sec:ubitect}.

\subsubsection{Apply Path Analysis}
\label{subsubsec:postcon_rule}

Following \S\ref{subsec:postcondi_work}, Figure \ref{fig:postcondi_use} presents a concert example of post-constraint guided path analysis. This case shows a simple initializer \(i(a)\) of the variable \(a\). Given an early return, the initialization in line 4 may not be executed. As such, the \textit{qualified postconditions} become contingent on the \textit{post-constraints} \(\mathcal{C}_{post}\). There are:

\squishlist
\item If the use of variable \texttt{a} is unconditional, \ie \(\mathcal{C}_{post}=\top\).
In this case, the variable \(a\) is labeled as \texttt{may\_init} given that the initialization \textit{may not} be reached.

In general, if all path constraints and outcomes of \texttt{must\_init} are \textit{disjoint from} \(\mathcal{C}_{post}\),
no path can be pruned out. We could also conclude \(a\) as \textit{may\_init}.
\item If the use of variable \(a\) is conditional with constraints, \ie \(\mathcal{C}_{post}\neq\top\), two cases emerge:
\begin{enumerate}
\item \(\mathcal{C}_{post}\) clashes with the constraints of the path (e.g., \texttt{some\_condi}), or
\item \(\mathcal{C}_{post}\) conflicts with the path outcome (e.g., \texttt{return -1}).
\end{enumerate}
In these instances, \(\mathcal{C}_{post}\) could be \texttt{some\_condi} or \texttt{func(...)==0} and we can designate \texttt{*a} as \texttt{must\_init}. 
\squishend

\cut{
\vspace{3pt}
\noindent
While there are a few other cases we have not considered for postconditions throughout the Linux kernel, the above rules cover most of the scenarios we have encountered. We highlight some interesting examples in \S\ref{subsec:excuse}. It's worth noting that the few-shots in-context learning methodology is extensible, making it easy to incorporate new rules and scenarios.
}

\begin{figure}
\begin{minipage}{.15\textwidth}
\centering
\begin{lstlisting}[language=c]
int func(int* a){  
  if(some_condi)
    return -1;
  *a = ... // init 
  return 0;
}
\end{lstlisting} 
\end{minipage}%
\begin{minipage}{.35\textwidth}
\centering
\vspace{-8pt}
\begin{tabular}{p{4.5cm}}

\text{\small \(\texttt{must\_init} =\emptyset\)  if:} \\
\text{\small \(\mathcal{C}_{post} = \top \)} or  \\
\text{\small\(\forall ps \in \{ \neg \texttt{some\_condi} \}: ps \perp \mathcal{C}_{post}\ \wedge \)} \\
\hspace{15pt}\text{\small \(\forall o \in \{ret \mapsto 0\}: o \perp \mathcal{C}_{post}\) } \\
\hline
\text{\small  \(\texttt{must\_init} =\{a\}\)  if:} \\
\text{\small \( (\neg \texttt{some\_condi}) \wedge \mathcal{C}_{post} \) or} \\
\text{\small\( (ret \mapsto 0) \wedge \mathcal{C}_{post} \) }\\
\end{tabular}

\end{minipage}

\caption{A sample case of initializer \texttt{func}, \texttt{*a} is \texttt{may\_init} or \texttt{must\_init} under different post-constraints. }
\label{fig:postcondi_use}
\end{figure}

\subsection{Design \#2: Progressive Prompt}
\label{subsec:iter_prompt}

\cut{
In the Linux kernel, it is common for each function to call many other functions further. Therefore, for the initializer summarizing task, if we do not provide
any further function definitions and ask LLMs to deliver a response instantly, they might fail due to a lack of knowledge.
However, if we put every relevant function definition at once, it will quickly exceed the limitation of context windows of LLMs.

To solve this dilemma, we propose to provide function definitions \textit{what LLMs need} selectively. As demonstrated in Figure \ref{fig:wf}, 
when LLMs analyze the initializer, the \textit{Progressive Prompt} design
fosters an ongoing interaction with the LLMs instead of soliciting an immediate response.
In this progressive exchange, 
we always prompt LLMs with: ``\textit{If you experience uncertainty due to insufficient function definitions, please indicate, and I will provide them}''. Upon receiving a request for additional information from LLM, we animatedly extract the necessary information from the source code and provide them to LLMs, enabling them to reevaluate and generate an improved response.
This back-and-forth interaction continues until the LLM garners sufficient information to deduce the initializer's behavior or until we exhaust the available information. If we cannot provide further information, for example, we can not find the requested function in the Linux source; we prompt it to continue analysis conservatively.
} 

The Linux kernel has an extremely large codebase. Summarizing an initializer using LLMs without providing any supplementary function definitions can result in incomplete or erroneous responses. On the other hand, flooding the LLM with every relevant function definition upfront risks exceeding their context window limitations.

To address this dilemma, we choose to progressively provide function definitions as needed. 
Illustrated in Figure \ref{fig:wf}, this approach, which we refer to as \textit{Progressive Prompt}, fosters a dynamic interaction with the LLM rather than expecting a response in one shot.
Throughout this iterative exchange, we consistently prompt the LLM: \textit{``If you encounter uncertainty due to a lack of function definitions, please signal your need, and I'll supply them''}. 
Should the LLM need more information, \work will promptly extract the relevant details on demand from the source code and provide it to the LLM \textit{automatically}, enabling it to reassess and generate a more accurate response. 

Specifically, We teach the LLM to ask for more information with a specific format:
\begin{lstlisting}[numbers=none]
    [{"type":"function_def", "name":"some_func" }]
\end{lstlisting}

Subsequently, \work scans this format in the LLM's response. For each requested function definition,
\work supplies its corresponding code along with comments extracted from the Linux source code.
Though GPT-4 may seek other types of information beyond function definitions (\eg struct definitions), we currently limit our support to requests pertaining to function definitions.

The iterative process continues until either the LLM no longer requests additional information, or \work cannot supply the requested details. In certain situations where \work is unable to provide more information (\eg the definition of an indirect call), \work will still prompt the LLM to proceed with the analysis. In these instances, the LLM is encouraged to infer the behavior based on the available data and its inherent knowledge, thereby facilitating continued analysis even when not all information is directly accessible.

\subsection{Design \#3: Task Decomposition}
\label{subsec:multi-step}

We systematically apply the principle of task decomposition, a vital element of our design process. This concept is incorporated primarily in two distinct ways.

\vspace{3pt}
\noindent
\textbf{Multistage Problem Solving.} 
\cut{
First, as Figure \ref{fig:wf} outlined, 
\work employs two separate \textit{conversations} to finish the task. Each conversation (\ie independent sessions) contains
several turns of \textit{prompts and responses}. \work first
extracts the initializer and its \textit{post-constraints} (subtasks 1 and 2) in Convo.1, 
and then summarizes the postconditions (subtask 3) based on the \textit{post-constraints} in Convo.2. 
Without task decomposition, we will combine all three subtasks into a single conversation, and prompt it to finish all three subtasks at once.
We also evaluate this option in \S\ref{subsec:expr_comp}.
}
As illustrated in Figure \ref{fig:wf}, we employ a two-conversation approach to complete the task. Each conversation, essentially consists of multiple iterations of prompts and responses. The first conversation (Convo.1) is dedicated to extracting the initializer and its associated post-constraints (subtasks 1 and 2), while the second conversation (Convo.2) focuses on summarizing the function (subtask 3) based on the previously identified post-constraints. This division allows a more manageable and effective way of achieving the task, compared to combining all three subtasks into a single conversation. The efficacy of this task decomposition approach is further evaluated in \S\ref{subsec:expr_comp}.

\vspace{3pt}
\noindent
\textbf{Thinking in English.} 
Our workflow necessitates a structured output, such as a JSON format, for automation. However, we observe that LLMs often produce suboptimal results when directly prompted to output in this format. As LLMs build responses incrementally, word-by-word, based on preceding outputs \cite{vaswani_attention_2017}, direct prompts to output JSON may interrupt their thought progression. This emphasizes the importance of initially soliciting responses in natural language to ensure comprehensive and effective reasoning. Consequently, we instruct the LLM to first articulate their thought processes in English, followed by a subsequent prompt to transform their response into a JSON summary.


\subsection{Design \#4: Self-Validation}
\label{subsec:self_refine}

\cut{
LLMs sometimes exhibit erratic or inconsistent behaviors, especially in complex scenarios that involve intricate logic. For example, for a initializer 
with postcondition \texttt{must\_init} if \(ret \mapsto 0\), the LLM 
sometimes might ignore the post-constraint and believe it is
\texttt{may\_init} even with the exact 
post-constraint \(ret \mapsto 0\).
On the other hand, LLM might pretend 
non-exist post-constraint and mistakenly conclude
a \texttt{may\_init} case to \texttt{must\_init}.
Namely, LLMs brings both false positives and false negatives in detecting bugs.
}

At times, LLMs can display unpredictable or inconsistent behaviors, particularly in complex scenarios involving detailed logical constructs. Consider a case where an initializer carries the postcondition \texttt{must\_init} if \(ret \mapsto 0\). LLMs may still mistakenly assume it to be \texttt{may\_init}, despite the explicit presence of the post-constraint \(ret \mapsto 0\).

Conversely, an LLM might erroneously interpret a non-existent post-constraint and incorrectly infer a \texttt{may\_init} case as \texttt{must\_init}. 
This phenomenon is known as \textit{hallucination}.
Essentially, the hallucination can lead to both false positives and false negatives in bug detection, thereby affecting accuracy and reliability.

In addition to task decomposition,  we also introduce the concept of \textit{self-validation} to enhance reliability.
Before the LLM reaches its final conclusion,
this method reinforces specific rules, allowing the LLM to reassess their previous responses for adherence and make necessary corrections. We observed that this practice yields better results. We evaluate the effect of self-validation in \S\ref{sec:comparison}.

As seen in Figure \ref{fig:wf}, we employ self-validation in both conversations. 
By prompting a list of  \textit{correct} properties that we expect, LLMs can verify and correct their results by themselves automatically.
\subsection{Additional Prompting Strategies}
\label{subsec:other_prompt}

In order to further optimize the efficacy of our model, we have incorporated several additional strategies into our prompt design:

\squishlist

\item \textbf{Chain-of-Thought.}
Leveraging the Chain-of-Thought (CoT) approach, we encourage the LLMs to engage in stepwise reasoning, using the phrase \textit{``think step by step''}. This not only helps generate longer, comprehensive responses, but it also provides intermediate results at each juncture of the thought process. Previous studies suggest the CoT approach considerably enhances the LLMs' reasoning capabilities~\cite{chen_when_2023}. We incorporate the CoT strategy into every prompt.

\item \textbf{Source Code Analysis.}
Rather than analyzing abstract representations, we opt to focus our attention directly on the functions within the source code. This approach not only economizes on token use compared to LLVM IR, but also allows the model to leverage the semantic richness of variable names and other programming constructs to conduct a more nuanced analysis.


\squishend


\cut{
There are many interesting details in building effective prompts.
For example, prompting LLM output with \textit{conditions} of each \texttt{may\_init} (\ie in what cases it will initialize) improves its performance. Moreover, sometimes replace a word by its synonym can impact the results. 
For example, if we tell LLM with \textit{`don't do something'} --
it sometimes understands exactly the opposite -- \textit{`do something'}.
Considering they do not affect the overall approach, interested readers can discover these details from our open-sourced detailed prompts implementation and experiments.
}

\cut{
\vspace{3pt}
\noindent
Designing effective prompts involves mastering many intriguing nuances. For instance, enhancing the LLM's output by specifying the \textit{conditions} under which each \texttt{may\_init} (i.e., conditions in which initialization will occur) can boost its performance. 
Furthermore, equally impactful can be the strategic substitution of words with their synonyms. A striking example is the paradoxical interpretation of negations by the LLM. If prompted with a command like \textit{`don't do something'}, the LLM occasionally comprehends it as \textit{`do something'}, the exact reverse of the intended instruction. 
\haonan{I personally like it, but can also remove}
}

\vspace{3pt}
\noindent
There are still some interesting details in designing an effective prompt
but due to space constraints and without changing the overall strategy, we will not list them all. Readers intrigued can delve into the intricacies of our open-sourced prompt\footnote{\href{https://sites.google.com/view/llift-open/prompt}{https://sites.google.com/view/llift-open/prompt}} design and experimental implementations to gain a deeper understanding.
\section{Implementation}

We implement the prototype of \work based on OpenAI's API~\cite{openai_2022_introducing_2022}
(\ie gpt-4-0613).
We describe some implementation details in the following aspects:

\vspace{3pt}
\noindent
\textbf{Interaction with LLMs.} \work's interaction with LLMs is managed by a simple agent developed in Python, containing roughly 1,000 lines of code. In addition, it uses seven prompts, which altogether constitute about 2,000 tokens in two conversations.
All interactions are \textit{fully automated} via APIs of OpenAI. 
Besides sending prompts and waiting for responses, our agent also 1) interacts with LLMs according to the progressive prompt design, 2) locates function definitions within the Linux source code, and 3) processes responses from LLMs, then receives and stores to a database.

\vspace{3pt}
\noindent
\textbf{Hyper-Parameters.} There are several hyper-parameters in calling the APIs provided by OpenAI. We choose \texttt{max\_token} and \texttt{temperature} to 1,024 and 1.0, respectively. \texttt{max\_token} controls the output length; since LLMs always predict the next words by the previous output, the longer output can benefit and allow its reasoning. However, too many tokens will exhaust the context window quickly, so we pick 1024 as a reasonable balance.

The temperature controls the randomness and also the ability to reason. 
Intuitively, we want the analysis to be as non-random as possible and reduce the temperature (it can take a value between 0 and 2 for GPT models);
however, an overly low temperature can result in repetitive or overly simplistic responses. We set it to 1.0 (also the default of gpt-4-0613), which allows for higher-quality responses, and use strategies such as self-validation and majority voting to improve the consistency of responses.

\begin{table*}[]
    \centering
    \caption{True bugs identified by \work from Random-1000, analyzing in Linux v4.14}
    \label{tab:table_rq1}
\scalebox{0.9}{
\begin{tabular}{llllll}
\toprule
\textbf{Initializer}               & \textbf{Caller}                            & \textbf{File Path }                                      & \textbf{Variable}          & \textbf{Line} \\
\midrule
read\_reg                & get\_signal\_parameters              & drivers/media/dvb-frontends/stv0910.c      & tmp                & 504                                 \\
regmap\_read             & isc\_update\_profile                 & drivers/media/platform/atmel/atmel-isc.c   & sr                 & 664                                 \\
ep0\_read\_setup         & ep0\_handle\_setup                   & drivers/usb/mtu3/mtu3\_gadget\_ep0.c       & setup.bRequestType & 637                                 \\
regmap\_read             & mdio\_sc\_cfg\_reg\_write            & drivers/net/ethernet/hisilicon/hns\_mdio.c & reg\_value         & 169                                 \\
bcm3510\_do\_hab\_cmd    & bcm3510\_check\_firmware\_version    & drivers/media/dvb-frontends/bcm3510.c      & ver.demod\_version & 666                                 \\
readCapabilityRid        & airo\_get\_range                     & drivers/net/wireless/cisco/airo.c          & cap\_rid.softCap   & 6936                                \\
e1e\_rphy                & \_\_e1000\_resume                    & drivers/net/ethernet/intel/e1000e/netdev.c & phy\_data          & 6580                                \\
pci\_read\_config\_dword & adm8211\_probe                       & drivers/net/wireless/admtek/adm8211.c      & reg                & 1814                                \\
lan78xx\_read\_reg       & lan78xx\_write\_raw\_otp             & drivers/net/usb/lan78xx.c                  & buf                & 873                                 \\
t1\_tpi\_read            & my3126\_phy\_reset                   & drivers/net/ethernet/chelsio/cxgb/my3126.c & val                & 193                                 \\
pci\_read\_config\_dword & quirk\_intel\_purley\_xeon\_ras\_cap & arch/x86/kernel/quirks.c                   & capid0             & 562                                 \\
ata\_timing\_compute     & opti82c46x\_set\_piomode             & drivers/ata/pata\_legacy.c                 & \&tp               & 564                                 \\
pt\_completion           & pt\_req\_sense                       & drivers/block/paride/pt.c                  & buf                & 368                                \\
\bottomrule
\end{tabular}
}

\end{table*}

\section{Evaluation}
\label{sec:eval}

Our evaluation aims to address the following research questions.

\squishlist
\item \textbf{RQ1 (Precision):} How accurately is \work able to identify bugs?
\item \textbf{RQ2 (Recall):} Is there a possibility for \work to miss real bugs?
\item \textbf{RQ3 (Comparison):} How does the performance of individual components within \work compare to that of the final design?
\item \textbf{RQ4 (Model Versatility):} How does \work perform when applied to LLMs other than GPT-4?
\squishend

\vspace{3pt}
\noindent

We evaluate RQ1 to RQ3 in GPT-4, under API from OpenAI with version gpt4-0613.
For RQ4, we also test GPT-3.5 with version gpt-3.5-turbo-0613 and Claude 2 additionally for comparison.

\subsection{Dataset}


Our experiment data, sourced from UBITect, includes all potential bugs labeled by its static analysis stage but experienced timeout or memory exhaustion during its symbolic execution stage. 
Overall, UBITect's static analysis stage produced 140,000 potential bugs, with symbolic execution able to process only 60\%, leaving 53,000 cases unattended, which means that these cases are generally difficult for static analysis or symbolic execution to decide
We craft the following dataset from 53,000 cases to evaluate \work:

~\noindent(1) \textbf{Random-1000.} We randomly chose 1,000 from the 53,000 cases for testing. However, there are 182 cases where there are no initializers, which are automatically recognized and filtered (see \S\ref{sec:problem}). The remaining 818 cases are used in evaluating precision, \ie the ratio of true positives to false positives. 

~\noindent(2) \textbf{Bug-50.} This dataset comprises the 52 confirmed UBI bugs previously identified by UBITect. It is used as ground truth for assessing recall by verifying if any true bugs were overlooked.

~\noindent(3) \textbf{Cmp-40.} This dataset comprises 27 negative and 13 positive cases selected from the Random-1000. We utilize this dataset to illustrate which of our design strategies contributed most to the outcome of our solution.

\PP{Turns and Conversations.}
Due to the progressive prompt, each case may require different turns (pairs of a prompt and a response). In Random-1000, the average number of turns is 2.78, with a max of 8 and a variance of 1.20.

\PP{Cost.} On average, it costs 7,000 tokens in GPT-4 to analyze each potential bug.



\cut{
\begin{figure}
\begin{minted}[xleftmargin=10pt, linenos, fontsize=\footnotesize]{c}
int ath10k_pci_diag_write_mem(..., int nbytes){
 while (ath10k_ce_completed_recv_next_nolock(..., 
           &completed_nbytes) != 0) {
    mdelay(1);
    if (i++ > DIAG_ACCESS_CE_TIMEOUT_MS) {
      ret = -EBUSY;
      goto done;
    }
  }
  //use of "completed_nbytes"
  ...

}
int ath10k_ce_completed_recv_next_nolock(..., unsigned int *nbytesp){
 ...
 nbytes = __le16_to_cpu(sdesc.nbytes);
 if (nbytes == 0) {
   return -EIO;
 }
 desc->nbytes = 0; 
 /* Return data from completed destination descriptor */
 *nbytesp = nbytes;
   ...
 return 0;
}
\end{minted}
\caption{Code snippet of \texttt{ath10k\_...\_nolock} and its usecase, derived from \texttt{net/wireless/ath/ath10k/ce.c} }
\label{fig:case_while}  
\end{figure}
}

\subsection{RQ1: Precision}
\label{subsec:expr_precision}



\work reports 26 positives among the Random-1000 dataset, where half of them are true bugs based on our manual inspection. This represents a precision of 50\%.
In keeping with UBITect and we focus on the analysis of Linux v4.14,  12 of the bugs still exist in the latest Linux kernel.
We are in the process of reporting the 12 bugs to the Linux community.
So far, we have submitted patches for 4 bugs and received confirmation that they are true bugs.


\vspace{3pt}
\noindent
\textbf{Imprecise and Failed Cases.} 
 Despite the effectiveness of \work, there are instances where it does not yield precise results, resulting in 13 false positives by mistakenly classifying \texttt{must\_init} cases as \texttt{may\_init}. Upon a careful examination of these cases, we attribute the imprecision to a variety of factors, which we discuss in detail in \S\ref{subsec:excuse}.
 Briefly, we give a breakdown of them here: \textit{Incomplete constraint extraction} (4 cases), \textit{Information gaps in UBITect} (5 cases), \textit{Variable reuse} (1 case), \textit{Indirect call} (1 case), and \textit{Additional constraints} (1 case). Additionally, there is one false positive caused by inconsistent output (i.e., two false positives in three runs). 
 Four cases exceed the maximum context length while exploring deeper functions in the progressive prompt. 

\find{
\textbf{Takeaway 1.} \work Can effectively summarize initializer behavior 
and discover new bugs with high precision (50\%).
}

\cut{
\begin{figure}
    \centering
    \includegraphics[width=0.45\textwidth]{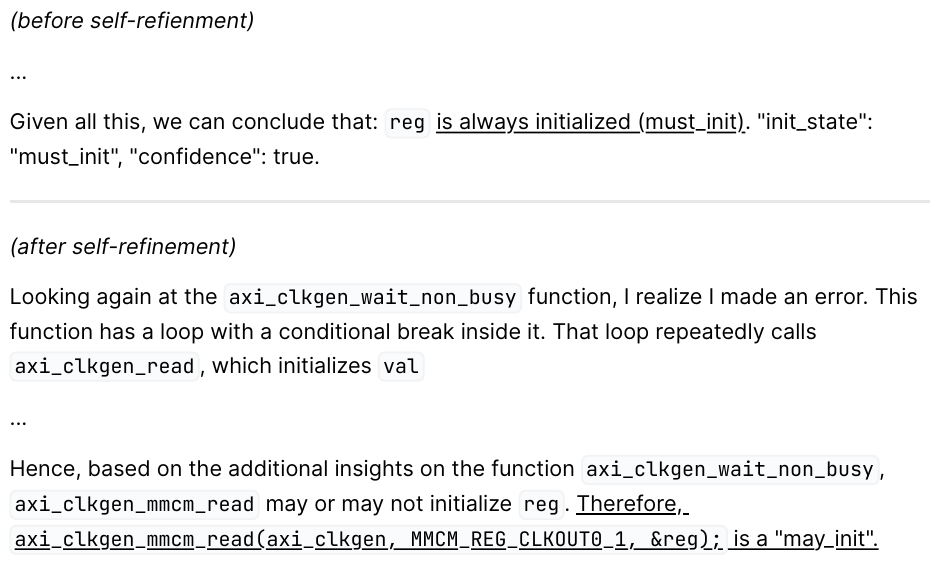}
    \caption{self-validation of analyzing case \texttt{axi\_clkgen\_recalc\_rate}, it concludes an incorrect answer
    \texttt{must\_init} at first, and then corrects it to \texttt{may\_init} after the self-validation.}
    \label{fig:con_sr}
\end{figure}
}

\subsection{RQ2: Recall Estimate}
\label{subsec:expr_recall}

Conceptually, the core optimization (post-constraint guided path analysis) of \work is sound, 
and we also prompt a series of rules to let LLMs tend to respond ``\texttt{may\_init} when uncertain.
We expect \work would not reject true bugs or with a high recall.

We sample 300 negative cases
from Random-1000 in an effort to see whether we will miss any true bugs. 
We confirm that all are true negatives. Despite the limited data sampled, this result indicates that integrating GPT-4 into our implementation does not introduce apparent unsoundness. 

Further, we test \work on the Bug-50 dataset to see \textit{whether it will miss any bugs discovered by UBITect}.
\work has demonstrated full effectiveness in identifying all real bugs from Bug-50. This result, while encouraging, does not imply that \work is flawless. Detailed data analysis reveals that:
1) There remain some inconsistencies in 3\(\sim\)5 cases
occasionally, though they are mitigated by majority voting; and 2) 
all the bugs found by UBITect have trivial post-constraints \((\mathcal{C}_{post}=\top\)) and postcondition of \textit{may\_init} (\(\mathcal{P}_{qual}: \texttt{must\_init} \mapsto \emptyset\)). 
Hence, \work could identify them easily.
It is noteworthy that these cases are already those cases detectable by UBITect. 
Such cases tend to be simpler in nature and can be verified by symbolic execution in UBITect.

\find{
\textbf{Takeaway 2.} \work has proven effective in identifying UBI bugs, consistently detecting all known instances.
}

\begin{table}[t]
\centering
\caption{Performance evaluation of bug detection tool with progressive addition of design components: Post-Constraint Guided Path Analysis (PCA), Progressive Prompt (PP), Self-Validation (SV), and Task Decomposition (TD). (C) indicates the number of
\textit{C}onsistent cases.}
\scalebox{0.78}{
\begin{tabular}{@{}lcc|cccc@{}}
\toprule
\textbf{Combination} & \textbf{TN(C)} & \textbf{TP(C)} & \textbf{Precision} & \textbf{Recall} & \textbf{Accuracy} & \textbf{F1 Score} \\ \midrule
Simple Prompt & 12(9) & 2(1) & 0.12& 0.15& 0.35& 0.13 \\
PCA & 13(9) & 5(1) & 0.26 &	0.38 & 0.45 &	0.31 \\
PCA+PP & 5(3) & 6(1) & 0.21& 0.46& 0.28& 0.29 \\
PCA+PP+SV & 5(2) & 11(8) & 0.33& 0.85& 0.40& 0.48 \\
PCA+PP+TD & 22(14) & 6(4) & 0.55& 0.46& 0.70& 0.50 \\ 
{PCA+PP+SV+TD} & 25(17) & 13(12) & 0.87& 1.00& 0.95& 0.93 \\ \midrule
\textit{Oracle} & 27(27) & 13(13) & - & - & - & - \\
\bottomrule
\end{tabular}
}
\label{tab:compare}
\end{table}

\subsection{RQ3: Contributions of Design Strategies}
\label{sec:comparison}

In our effort to delineate the contributions of distinct design strategies to the final results, we undertook an evaluative exercise against the Cmp-40 dataset, employing varying configurations of our solution, each entailing a unique combination of our proposed strategies. As illustrated in Table \ref{tab:compare}, the strategies under consideration encompass Post-constraint Analysis (\textit{PCA}), Progressive Prompt (\textit{PP}), Self-Validation (\textit{SV}), and Task Decomposition (\textit{TD}). The findings underscore an overall trend of enhanced performance with the integration of additional design strategies.

In this study, the \textit{Baseline} corresponds to a straightforward prompt, \textit{"check this code to determine if there are any UBI bugs"}, a strategy that has been found to be rather insufficient for discovering new vulnerabilities, as corroborated by past studies \cite{openai_2023_gpt_4, ma_scope_2023, tian_is_2023}, reflecting a modest recall rate of 0.15 and a precision of 0.12.


Incorporating PCA offers a notable enhancement, enabling the LLM to uncover a wider array of vulnerabilities. As shown in Table \ref{tab:compare}, there is a substantial improvement in recall in comparison to the baseline, an anticipated outcome considering PCA's pivotal role in our solution. However, solely relying on this strategy still leaves a lot of room for optimization.

The influence of Progressive Prompt (\textit{PP}) on the results is quite intriguing. While its impact appears to lower precision initially, the introduction of task decomposition and self-validation in conjunction with PP reveals a substantial boost in performance. Without PP, the LLM is restricted to deducing the function behavior merely based on the function context's semantics without further code analysis. Even though this approach can be effective in a range of situations, it confines the reasoning ability to the information available in its training data. By checking the detailed
conversation, we notice the omission of TD or SV tends to result in the LLM neglecting the post-constraints, subsequently leading to errors.

Beyond influencing precision and recall, Task Decomposition (\textit{TD}) and Self-Validation (\textit{SV}) also play a crucial role in enhancing \textit{consistency}. In this context, a result is deemed \textit{consistent} if the LLM yields the same outcome across its initial two runs. A comparison between our comprehensive final design encompassing all components, and the designs lacking TD and SV, respectively, reveals that both TD and SV notably augment the number of consistent results, and deliver 17 and 23 consistent results in its negative and positive results, respectively, underscoring their importance in ensuring reliable and consistent outcomes.

Finally, TD also holds significance in terms of conserving tokens. 
In our evaluation phase, we identified two instances within the PCA+PP and PCA+PP+SV configurations where the token count surpassed the limitations set by GPT-4. However, this constraint was not breached in any case when TD was incorporated. 

\find{
\textbf{Takeaway 3.} All of \work's design strategies contributed to the positive results. 
}

\begin{table}[]
    \centering
    \caption{Comparison of different LLMs on real bugs, from a subset of Bug-50}
\scalebox{1.0}{
\begin{tabular}{lccccc}
\toprule
\multirow{2}{*}{\textbf{Caller}}            & \multicolumn{2}{c}{\textbf{GPT}} & \multirow{2}{*}{\textbf{Claude2}}    & \multirow{2}{*}{\textbf{Bard}}   \\
                                    &   \textbf{4}  & \textbf{3.5} &  &  \\
\midrule
\text{hpet\_msi\_resume}           &  \ding{51} &  \ding{51} &  \ding{51}  &  \ding{55}          \\
\text{ctrl\_cx2341x\_getv4lflags}  &  \ding{51} &  \ding{51} &  \ding{55}  &  \ding{55}          \\
\text{axi\_clkgen\_recalc\_rate}   &  \ding{51} &  \ding{51} &  \ding{51}  &  \ding{51}          \\
\text{max8907\_regulator\_probe}   &  \ding{51} &  \ding{51} &  \ding{51}  &  \ding{51}          \\
\text{ov5693\_detect}              &  \ding{51} &  \ding{51} &  \ding{55}  &  \ding{51}          \\
\text{iommu\_unmap\_page}          &  \ding{51} &  \ding{55} &  \ding{51}  &  \ding{55}          \\
\text{mt9m114\_detect}             &  \ding{51} &  \ding{51} &  \ding{51}  &  \ding{51}          \\
\text{ec\_read\_u8}                &  \ding{51} &  \ding{51} &  \ding{51}  &  \ding{51}          \\
\text{compress\_sliced\_buf}       &  \ding{51} &  \ding{51} &  \ding{55}  &  \ding{51}          \\
\bottomrule
\end{tabular}
}

\label{tab:res_cmp}
\end{table}

\subsection{RQ4: Alternative Models}
\label{subsec:expr_comp}


Table \ref{tab:res_cmp} provides a comprehensive view of the performance of our solution, \work, when implemented across an array of LLMs including GPT-4.0, GPT-3.5, Claude 2 \cite{anthropic_2023_claude}, and Bard \cite{krawczyk_bards_2023}. GPT-4 passes all tests, while GPT-3.5, Claude 2, and Bard exhibit recall rates of 89\%, 67\%, and 67\%, respectively. Despite the unparalleled performance of GPT-4, the other LLMs still produce substantial and competitive results, thereby indicating the wide applicability of our approaches.

It is imperative to note that not all design strategies in our toolbox are universally applicable across all language models. Bard and GPT-3.5, in particular, exhibit limited adaptability towards the progressive prompt and task decomposition strategies. Bard's interaction patterns suggest a preference for immediate response generation, leveraging its internal knowledge base rather than requesting additional function definitions, thereby hindering the effectiveness of the progressive prompt approach. Similarly, when task decomposition is implemented, these models often misinterpret or inaccurately collect post-constraints, subsequently compromising the results. To harness their maximum potential, we only apply the PCA design specifically (\ie without other design strategies) for GPT-3.5 and Bard.

Contrasting the GPT series, Bard and Claude 2 demonstrate less familiarity with the Linux kernel and are more prone to failures due to their unawareness of the \texttt{may\_init} possibility of initializers.


\find{
\textbf{Takeaway 4.} 
GPT-4 remains at the pinnacle of performance for \work, yet other LLMs can achieve promising results. 
}


\begin{figure}
\hspace{-15pt}
\includegraphics{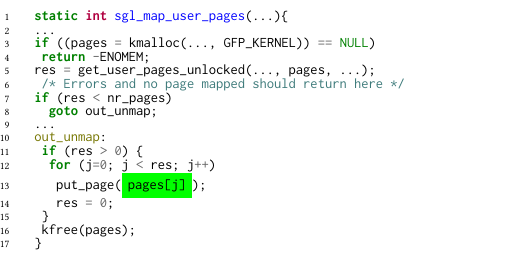}
    \caption{Case Study I (Loop and Index). Derived from \texttt{drivers/scsi/st.c} 
    }
    \label{fig:sgl_pages}
\end{figure}

\subsection{Case Study}
\label{subsec:case_study}
In this case study, we pick three interesting cases demonstrating the effectiveness of \work in analyzing function behaviors and detecting uninitialized variables. 
All these cases are undecided for the previous static analyzer, UBITect.
We put the complete conversations on an anonymous online page for reference\footnote{\href{https://sites.google.com/view/llift-open/case-studies}{https://sites.google.com/view/llift-open/case-studies}}.






\vspace{3pt}
\noindent
\textbf{Loop and Index.} Figure \ref{fig:sgl_pages} presents an intriguing case involving the variable \texttt{pages[j]}, which is reported by UBITect as used in Line 17 potentially without being initialized. Unfortunately, this case is a false positive which is hard to prune due to loops. Specifically, the initializer function \texttt{get\_user\_pages\_unlocked()}, which is responsible for mapping user space pages into the kernel space, initializes the \texttt{pages} array allocated in Line 3. If \texttt{get\_user\_pages\_unlocked()} is successfully executed, \texttt{pages[0]} through \texttt{pages[res-1]} pointers will be initialized to point to struct page instances. 

To summarize the behavior, \ie \texttt{must\_init} facts under conditions where the use is reachable,
we must first extract the \textit{post-constraints} that lead to the use of \texttt{pages}. Through interacting with ChatGPT, \work successfully extracts it:
\begin{lstlisting}[numbers=none]
{
    "initializer": "res = get_user_pages_unlocked(uaddr, nr_pages, pages, rw == READ ? FOLL_WRITE : 0)",
    "suspicious": ["pages[j]"],
    "postconstraint": "res < nr_pages && res > 0 && j < res",
}
\end{lstlisting}

After feeding the post-constraints to LLM, \work then successfully obtains the result:

\begin{lstlisting}[numbers=none]
{
    "ret": "success",
    "response": {
        "must_init": ["pages[j]"],
        "may_init": [],
    }
}
\end{lstlisting}

As we can see, GPT-4 exhibits impressive comprehension of this complex function. 
It perceives the variable \texttt{pages[j]} being used in a loop that iterates from \texttt{0} to \texttt{res-1}. This insight leads GPT-4 to correctly deduce that all elements in the \texttt{pages} array must be initialized, \ie they are \texttt{must\_init}. This example underscores GPT-4's proficiency in handling loop and even index sensitivity.

\begin{figure}
\hspace{-15pt}
\includegraphics{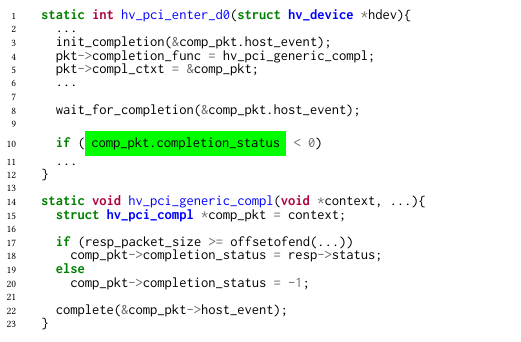}
    \caption{Case Study II (Concurrency and Indirect Call). Derived from \texttt{drivers/pci/host/pci-hyperv.c}
    }
    \label{fig:hv_pci}
\end{figure}

\vspace{3pt}
\noindent
\textbf{Concurrency and Callback.} Consider the case illustrated in Figure \ref{fig:hv_pci}. At first glance, UBITect flags Line 10 for potentially using the variable \texttt{comp\_pkt.completion\_status} before initialization. The function's body seemingly lacks any code that initializes it, leading UBITect to report it as a potential bug.  However, the mystery unravels when we examine \texttt{hv\_pci\_generic\_compl()}, the actual initializer function assigned to \texttt{pkt} in Line 4. The variable in question is indeed initialized, but intriguingly, its initializer emerges from a concurrent function instead of within its own thread. Here \texttt{wait\_for\_completion()} is a synchronization primitive that pauses the current thread and waits for the new thread (\ie \texttt{hv\_pci\_generic\_compl()}) to complete.
Despite this complexity, GPT-4 adeptly navigates the concurrency and callback handling, pinpointing the accurate initializer and outputting a precise result.

It is worth noting that we do not encode any knowledge about the Linux kernel synchronization primitives. 
\work prompts LLMs with  \textit{``The `initializer' must be the `actual' function that initializes the variable.''}
and then LLMs can automatically identify the function \texttt{hv\_pci\_generic\_compl()}
as the initializer of \texttt{comp\_pkt.completion\_status}.

\begin{figure}[]
\hspace{-15pt}
\includegraphics{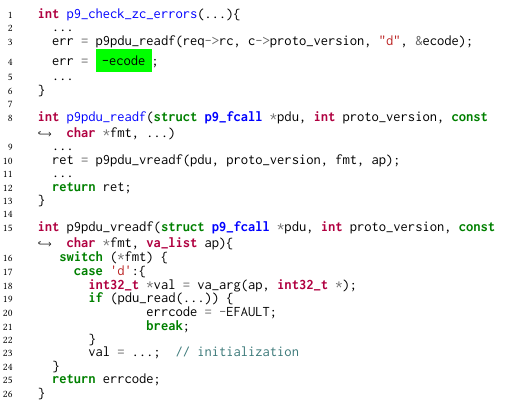}
    \caption{Case Study III (Unfamiliar Function), derived from \texttt{net/9p} 
    }
    \label{fig:p9_read}
\end{figure}

\vspace{3pt}
\noindent
\textbf{Unfamiliar Function.} 
As previously delineated in \S\ref{subsec:cap}, LLMs possess the inherent ability to recognize the semantics (\eg postconditions) of common functions like \texttt{sscanf()}. However, some argue that \textit{``the LLM simply learns everything from the internet and acts merely as a search engine''} \cite{chiang_chatgpt_2023}. This viewpoint is challenged by the case illustrated in Figure \ref{fig:p9_read}.

The case presents an intriguing real-world bug. The function \texttt{p9pdu\_readf()} mirrors \texttt{sscanf()} in structure, yet lacks a check of its return value, leaving the parameter \texttt{ecode} at risk of being uninitialized, i.e., if \texttt{pdu\_read()} returns non-zero in line 19 (thus ``break'' early).
Notably, unlike \texttt{sscanf()}, where GPT-4 can provide a precise summary of the function without asking for its definition,
it does request the function definition of \texttt{p9pdu\_readf()}, as it is not as ubiquitous as \texttt{sscanf()}.  

Furthermore, our solution not only produces the correct outcome for this particular case but also pinpoints that \texttt{ecode} could be initialized when \texttt{p9pdu\_readf()} returns 0, demonstrating the efficacy of \work for unfamiliar cases. The result is as follows:


\begin{lstlisting}[numbers=none]
{
    "initializer": 
        "err = p9pdu_readf(req->rc, c->proto_version, 'd', &ecode)",
    "suspicious": ["ecode"],
    "postconstraint": null,
    "response": {
        "must_init": [],
        "may_init": [{
                "name": "ecode",
                "condition": "p9pdu_readf returns 0"
        }]
    }
}
\end{lstlisting}


\vspace{-7pt}
\subsection{Reason for Imprecision}
\label{subsec:excuse}


Despite \work achieving a precision of 50\% in real-world applications, 
the precision can still be improved in the future. Some 
can be solved with better prompts or better integration with static analysis.

\vspace{3pt}
\noindent
\textbf{Challenges in Constraint Extraction.} Beyond the four primary code patterns we addressed in \S\ref{subsec:postcondi}, there exist additional forms of post-constraints. For instance, during error handling, the checks for failures may involve another function or macro. This problem can be addressed by either more examples during prompts (in-context learning), or lightweight program analysis (\eg path exploration in symbolic execution to collect the post-constraints). 

\cut{
\vspace{3pt}
\noindent
\textbf{Insufficient Field Sensitivity.} LLMs can distinguish between different fields of a structure; furthermore, they can even identify
some particular cases such as \texttt{container\_of}. 
However, UBITect does not point out the field name by default.
With more dedicated engineering work in extracting field information from UBITect, the field sensitivity of \work can be substantially improved. 

\vspace{3pt}
\noindent
\textbf{Propagation of Uninitialization} In UBITect, if an initialized variable \textit{a} is \textbf{propagated} to another variable \textit{b}, it will correctly track \textit{b} as uninitialized. Therefore, if an initializer operates on \textit{b} instead of \textit{a}, and the subsequent use is also on \textit{b},
it will attribute the bug variable \textit{a}. 
Unfortunately, due to the lack of interface exposing the relationship between \textit{a} and \textit{b}, \work is aware of \textit{a} being the suspicious variable and fails to identify any initializer for it. This issue can be solved by better collaboration with UBITect.
}

\vspace{3pt}
\noindent
\textbf{Information Gaps in UBITect.} For instance, UBITect does not provide explicit field names within a structure when a specific field is in use. This information gap can result in \work lacking precision in its analysis. Additionally, UBITect only reports the variable utilized, not necessarily the same variable passed to an initializer. For example, consider an uninitialized variable \texttt{a} passed to an initializer, which is then assigned to variable \texttt{b} for usage. In such a scenario, \work may fail to identify the initializer due to this incomplete information correctly. These challenges, primarily due to the interface design in UBITect, can be addressed with focused engineering efforts to enrich the output information from UBITect.


\cut{
could lead another variable \textit{X} becoming uninitialized, if there's any assignment to \textit{X} like: \( X=E, \, v\in E \). We do not consider the propagation because we treat \(X=E\) as the use of \textit{v}, so this bug can be found in another warning. However, UBITect considers propagation and only outputs ``critical'' ones heuristically.

\work filters out most propagation cases automatically, because there it cannot find an initializer. 
A complete solution relies on letting UBITect output all potential bugs without heuristic selection,
and therefore we \haonan{...}
}

\vspace{3pt}
\noindent
\textbf{Variable Reuse.} 
Varaible reuse is an interesting problem of LLM. In general, LLM usually confuses different variables in different scopes (\eg different function calls).
For example, if the suspicious variable is \texttt{ret} and passed as a argument to its 
initializer (say, \texttt{func(\&ret)}) and there is another stack variable
defined in \texttt{func} also called \texttt{ret}, LLM will confuse them. 
Explicitly prompting and teaching LLM to note the difference does not appear to work. One solution is to leverage a simple static analysis to normalize the source code to ensure each variable has a unique name.

\vspace{3pt}
\noindent
\textbf{Indirect Call.} 
As mentioned \S\ref{subsec:iter_prompt}, \work follows a simple but imprecise strategy to handle indirect calls. 
Theoretically, existing static analysis tools, such as MLTA \cite{lu_where_2019}, can give possible targets for indirect calls. However, 
each indirect call may have multiple
possible targets and  dramatically increase the token usage. 
We leave the exploration of such an exhaustive strategy for future work.
 \work may benefit from a more precise indirect call resolution.


\vspace{3pt}
\noindent
\textbf{Additional Constraints.} There are many variables whose values are determined outside of the function we analyze, \eg preconditions capturing constraints from the outer caller. Since our analysis is fundamentally under-constrained, this can lead \work to incorrectly determine a \texttt{must\_init} case to be \texttt{may\_init}. 
Mitigating this imprecision relies on further analysis to provide more information.




\section{Discussion and Future Work}
\label{sec:discuss}

\vspace{3pt}
\noindent
\textbf{Post-Constraint Analysis.} Our approach prioritizes post-constraints
over other constraints, such as preconditions. By focusing on the post-constraints, we enhance the precision and scalability significantly. 
Importantly, our utilization of large language models in program analysis
suggests strong abilities in summarizing complex function behaviors involving loops, a classic hurdle in program analysis.

\vspace{3pt}
\noindent
\textbf{Better Integration with Static Analysis.} Our work presents
opportunities for greater integration and synergy with static analysis methods.
Currently, our proposed solution operates largely independently of the static analysis methods, taking only inputs from static analysis initially.
Looking into the future, we can consider integrating static analysis and LLMs in a holistic workflow.
For example, this could involve selectively utilizing LLM as an assistant to overcome certain hurdles encountered by static analysis,
\eg difficulty in scaling up the analysis or summarizing loop invariants.
In turn, further static analysis based on these findings can provide insights to refine the queries to the LLM. This iterative process
could enable a more thorough and accurate analysis of complex cases.
We believe such a more integrated approach is a very promising future direction.

\cut{
For instance, preprocessing steps with static analysis could include tasks such
as variable normalization and conversion to SSA form. This can 
mitigate the confusion of variable reuse and varieties of postconditions.

On the flip side, the results from our LLM-based
approach could also be fed back into the static analysis process for further
examination. While a summary of \texttt{may\_init} and \texttt{must\_init} is
usually sufficient for detecting UBI bug, more complex bugs might
necessitate a feedback structure. This could involve utilizing the results from
the LLM, performing further static analysis based on these findings, and then
using these insights to refine the queries to the LLM. This iterative process
could enable a more thorough and accurate analysis for complex cases,
demonstrating the potential benefits of a more integrated approach combining
LLMs with traditional static analysis techniques.
}

\vspace{3pt}
\noindent
\textbf{Deploying on Open-sourced LLMs.}
The reproducibility of \work could be potentially challenged, considering its dependency on GPT-4, a closed-source API subject to frequent updates. 
At the time of writing, Meta introduced Llama 2, an open-source language model with capabilities rivaling GPT-3.5. Our initial assessments suggest that Llama 2 can understand our instructions and appears well-suited to support \work. The open-source nature of Llama 2 provides us with opportunities to deploy and refine the model further. We plan to leverage these prospects in future studies.

\section{Related Work}

\textbf{Techniques of Utilizing LLMs.}
Wang \etal \cite{wang_voyager_2023} propose an embodied lifelong learning agent based on LLMs.
Pallagani \etal \cite{pallagani_understanding_2023} explores the capabilities of LLMs for automated planning.
 Weng \cite{weng2023prompt}  summarizes recent work in building
an autonomous agent based on LLMs and proposes two important components for planning: \textit{Task Decomposition} and \textit{Self-reflection}, which
are similar to the design of \work.
Beyond dividing tasks into small pieces, task decomposition techniques also include some universal strategies such as
Chain-of-thought \cite{wei_chain--thought_2023} and Tree-of-thought \cite{yao_tree_2023}.
The general strategy of self-reflection has been used in several flavors: ReAct \cite{yao2023react}, Reflexion \cite{shinn_reflexion_2023} and Chain of Hindsight \cite{liu_chain_2023}.
Despite the similarity in name, self-reflection is fundamentally \textit{different} from self-validation in \work where the former focuses on using external sources to provide feedback to their models. Huang \etal \cite{huang_large_2022} let an LLM self-improve its reasoning without supervised data by asking the LLM to lay out different possible results.


\vspace{3pt}
\noindent
\textbf{LLMs for Program Analysis.} 
Ma \etal \cite{ma_scope_2023} and Sun \etal \cite{sun_automatic_2023}  explore the capabilities of LLMs when performing various program analysis tasks such as control flow graph construction, call graph analysis, and code summarization. They conclude that while LLMs can comprehend basic code syntax, they are somewhat limited in performing more sophisticated analyses such as pointer analysis and code behavior summarization.
In contrast to their findings, our research with \work has yielded encouraging results. We conjecture that this might be due to several reasons: (1) benchmark selection, \ie Linux kernel vs. others.
(2) Prompt designs. (3) GPT-3.5 vs. GPT-4.0 -- prior work only evaluated the results using only GPT-3.5.
Pei \etal \cite{pei_can_2023} use LLMs to reason about loop invariants with decent performance. In contrast, \work leverages LLMs for a variety of tasks (including program behavior summarization) and integrates them successfully into a static analysis pipeline.


\vspace{3pt}
\noindent \textbf{\acp{LLM} for Software Engineering}. 
Xia \etal \cite{xia_keep_2023} propose an automated conversation-driven
program repair tool using ChatGPT, achieving nearly 50\% success rate. 
Pearce \etal \cite{pearce_examining_2023} examine zero-shot vulnerability
repair using LLMs and found promise in synthetic and hand-crafted scenarios but
faced challenges in real-world examples.
Chen \etal \cite{chen_teaching_2023} teach LLMs to debug its own predicted program to increase its correctness, but only
performs on relatively simple programs.
Lemieux \etal \cite{lemieux_codamosa_2023} leverages LLM to generate
tests for uncovered functions when the search-based approach got coverage stalled. 
Feng and Chen \cite{feng_prompting_2023} use LLM to replay Android bug automatedly.
Recently, LangChain proposed LangSimith \cite{langchain_2023_announcing_2023}, a LLM-powered platform
for debugging, testing, and evaluating. 
These diverse applications underline the vast potential of LLMs in software engineering. \work complements these efforts by demonstrating the efficacy of LLMs in bug finding in the real world.


\section{Conclusion}




This work presents a novel approach that utilizes LLMs to aid static analysis using a completely automated agent.
By carefully considering the scope and designing the interactions with LLMs, our solution has yielded promising results.
We believe our effort only scratched the surface of the vast design space,
and hope our work will inspire future research in this exciting direction.


\cut{
In conclusion, paper presents a novel approach to detecting
use-before-initialization (UBI) bugs in the Linux kernel by employing large
language models (LLMs) such as ChatGPT. Our method involves generating precise
function summaries with the help of LLMs, which subsequently allows for more
accurate bug detection, reducing both false positives and false negatives.
}

\cut{
some questions copied from reviewer 2 for oakland poster, which may be helpful for future work, we can pick some:
How does the response for uncommon bugs look like? Did you observe imprecise responses for the bugs that are not very popular or appear in the codes of particular domains?
How do you address the issue of incorrect responses, and how do you measure the correctness of the responses? Does it require users to have sufficient domain knowledge regarding the bugs and the codes they are running the analysis on?
}

\bibliographystyle{ACM-Reference-Format}
\bibliography{main}


\end{document}